\documentstyle[preprint,aps]{revtex}

\begin{document}
\draft
\preprint{[\,preprint NA95b\,]}



\title{Composite-Fermion Analysis of the Double-Layer Fractional Quantum
Hall System}
\author{T. Nakajima$^1$ and H. Aoki$^2$}
\address{$^1$Semiconductor Research Laboratory, Mitsubishi Electric
Corporation, Amagasaki, Hyogo 661, Japan\\
$^2$Department of Physics, University of Tokyo, Hongo, Tokyo 113, Japan}
\date{June 28, 1995, cond-mat/9506124}

\maketitle

\begin{abstract}
Effect of interlayer tunneling in the double-layer fractional quantum Hall
system at the total Landau level filling of $\nu=1/m$ (\,$m$:\ odd
integer\,) is analyzed with the composite-fermion approach in which the
flux attachment is directly applied to the electron-electron interaction. A
comparison with a numerical result indicates that the vertically coupled
Laughlin liquids may be regarded as a system of composite fermions with
{\em reduced} \,interparticle interactions and {\em unchanged} \,interlayer
tunneling, which makes the quantum-Hall regime, identified by a gap in the
pseudospin-wave excitation mode, wider as $\nu$ becomes $1/3,1/5,\ldots$.
\end{abstract}




\newpage

\narrowtext

The fractional quantum Hall (FQH) state is a strongly-correlated quantum
liquid realized in a two-dimensional electron system in strong magnetic
fields \cite{qhe}.
Recently much attention is focused on what happens when two such liquids
are coupled face-to-face as realized in quantum wells \cite{dqw,suen}, and
the layer degrees of freedom in such double-layer systems are often
described by a pseudospin.

Although a (spin-polarized) double-layer system may at first seem analogous
to a single-layer system of spin 1/2 electrons from the analogy between
pseudospin and real spin, an essential difference does indeed exist through
two factors:
One is the interlayer tunneling, which makes single-particle wavefunctions
split into symmetric and antisymmetric (SAS) ones about the center of the
double-layer structure, with an energy separation, $\Delta_{\rm SAS}$. The
second is the controllability of the intra- versus inter-layer
electron-electron interaction strengths by the layer separation, which
makes the symmetry degrade from SU(2) to U(1).

In particular, the double-layer FQH state for the total Landau level
filling of $\nu=1/m$\ (\,$m$: odd integer\,) is pseudospin-polarized in
analogy with an easy-plane ferromagnet \cite{yang}.
The two factors above are exactly responsible for the pseudospin polarization:
(i) The tunneling gap acts as a magnetic field along the $x$ axis for the
pseudospin, thereby pushing the electrons into the symmetric band.
(ii) The intra- and inter-layer repulsive electron correlations make the
ground state represented by Halperin's $\Psi _{mmn}$ \cite{halperin83}, an
extension of Laughlin's state to two fermion species, where the same power
for intra- and inter-layer correlations (\,{\it i.e.}, $m=n$\,), realized
unless $d$ is too large, represents a pseudospin-polarized
state\cite{exact}.

Thus the layer separation, $d$,
normalized by the magnetic length, $\ell \equiv \sqrt{c \hbar/eB}$, and the
strength of tunneling, $\Delta _{\rm SAS}$, normalized by the
Coulomb-interaction energy, $e^2/\epsilon \ell$ (\,$\epsilon$: dielectric
constant\,) are the relevant dimensionless parameters, and we can think of
a phase diagram on the parameter plane. For an {\it integer} $\nu=1$, the
pseudospin-polarized quantum Hall state evolves continuously from the
correlation-dominated (or `two-component') character (\,$\Psi_{111}$\,) to
the tunneling-dominated (or `one-component') character. In the single-mode
approximation (SMA) \cite{mgb,issp}, this crossover, with no intervening
non-QHE region, is described by a continuous rotation of the pseudomagnon
vacuum in a Bogoliubov transformation.
The continuity is consistent with an experimental result by Murphy {\em et
al} \cite{dqw}.

There, the boundary of the QHE region is identified from a finite gap in
the charge
excitations.
Since the ground state is pseudospin polarized, the low-lying excitation is
a Goldstone (\,pseudospin-wave\,) mode where the densities of two layers
fluctuate in a correlated manner \cite{mgb,issp,hfrpa}.
MacDonald {\em et al} \,have in fact obtained the QHE region for $\nu=1$
from the gap in the pseudospin-wave mode with the SMA \cite{mgb}. The
excitation spectrum for $\nu=1$ has also been examined with the
random-phase approximation (RPA) or Hartree-Fock method\cite{hfrpa}.

A natural question then is how we can extend the physics of the
double-layer system with the inter-layer tunneling to the {\it fractional}
$\nu=1/3,1/5,\ldots$.
In fact the SMA calculations of the excitation spectrum have only been done
in the absence of interlayer tunneling for $\Psi_{mmn}$ with $\nu=2/(m+n)$
\cite{sma}.
An intriguing way we propose here is to apply a recent way of looking at
the ordinary (\,single-layer\,) FQH system in terms of the
composite-fermion picture \cite{jain} or the Chern-Simons gauge field
theory \cite{cs}, to the fractional double-layer system.

The composite-fermion (CF) picture asserts that a quantum Hall liquid of
electrons in an external magnetic field $B$ corresponding to
$\nu=p/(2mp+1)$ (\,$m$,\,$p$\,: integers\,) is equivalent, in a mean-field
sense, to a liquid of composite fermions each carrying $2m$ flux quanta in
a magnetic field $B_{\rm eff} \equiv B-B_{\nu=1/2m}$ corresponding to an
integer $\nu = |p|$.
There is a mounting body of numerical \cite{devwu} and experimental
\cite{exper} evidence supporting this picture.

It is a highly nontrivial problem whether the CF picture can correctly
describe the excitations that is dominated by both interaction and
tunneling.
As for the excitation spectra, a Chern-Simons (CS) approach, in which the
fluctuation around the mean CS gauge field is treated with the RPA, has
been developed \cite{coll}, but there the effective mass of the particle
remains a quantity difficult to fix.
In fact the Chern-Simons RPA calculations for the double-layer system
\cite{bcsdl,fcsdl} are still some way from an accurate description of the
intra-Landau-level excitations.
In this context it may be desirable to manipulate the interaction, so that
we adopt here a particular CF picture due to the present authors
\cite{swcf} that directly plugs the flux-attachment transformation into the
electron-electron interaction (\,Haldane's pseudopotential\,), which
enables us to obtain parameter-free results for the pseudospin-wave
excitation.

 We can then identify the QHE region from the charge (\,pseudospin-wave\,)
excitation spectrum, calculated from such a composite-fermion picture, in
the $\nu=1/m$ (\,$m$: odd integer\,) double-layer FQH system of
spin-polarized electrons.
A comparison with the exact result for a finite system shows that the
picture indeed gives a quantitatively accurate long-distance physics. This
provides the first analytic calculation of the excitation spectrum for a
fractionally filled system with interlayer tunneling.

Let us now mention two issues inherent in the double-layer FQH system.
One is whether there is an essential difference between a lateral tunneling
of composite particles (\,in a side-by-side geometry\,) \cite{feng} and the
vertical tunneling (\,in a face-to-face geometry\,).
We shall show that the $\nu=1/m$ double-layer FQH system can be regarded as
a system of composite fermions with the {\em reduced} \,inter-particle
interaction while the tunneling, when vertical, is {\em unchanged}.
The increased relative importance of the tunneling makes the QHE region for
$\nu=1/m$ in the phase diagram, obtained here for the first time for
$m=3,5$, wider for larger $m$.

Secondly, in the absence of inter-layer tunneling, the pseudospin-polarized
state has a broken U(1) symmetry and may have a spontaneous interlayer
phase coherence \cite{yang}.
The Goldstone (\,pseudospin-wave\,) mode restoring this broken symmetry is
in fact gapless and $k$-linear in analogy with a magnon in an XXZ
ferromagnet.
An introduction of the interlayer tunneling enforces the U(1) symmetry to
break, thereby introducing a gap in the pseudospin-wave mode.
The issue is whether the gapless Goldstone mode in the absence of
interlayer tunneling will signify a Josephson-like effect \cite{bcsdl,wen}.

We shall also touch upon this problem.
Hereafter we ignore the real spin degrees of freedom or the finite
thickness of each layer for simplicity.

The CF picture for the pseudospin-wave is the following.
When we attach $(m-1)$ flux quanta from the external field to each electron
for an odd $m=\nu ^{-1}$, the relative angular momentum $n$ between
electrons translates into the relative angular momentum $n-(m-1)$ between
composite fermions \cite{halperin82}.
Since $B_{\rm eff}=B/m$ is thereby reduced from the bare field by a factor
of $1/m$, the magnetic length $\ell$ changes into
$\tilde{\ell}=\sqrt{m}\,\ell$ in a mean-field sense, while the number of
single-particle states per unit area, $1/2 \pi \ell ^2$, is also reduced by
a factor $1/m$.

For the motion within a layer, we can work with the spherical geometry to
make the relevant quantum number the angular momentum.
As one maps stereographically a flat system onto a spherical one, the
translational symmetry is translated into the rotational symmetry, where
the wavenumber $k$ relates to the total angular momentum $L$ as $k = L/R$
with $R$ being the radius of the sphere. When the total magnetic flux going
out of the sphere is $2S$ (\,an integer due to Dirac's condition\,) times
the flux quantum, the relation to $\nu$ is $2S = \nu^{-1}N - \delta$ with
$N$ being the number of electrons and $\delta$ an integer.

The transformation into the CF picture is then given by
\begin{eqnarray}
2\tilde{S} = 2S/m &=& N-1, \nonumber \\
\tilde{V}_{2\tilde{S}-[n-(m-1)]}^{\sigma \sigma^{'}}/\ \frac{\
e^2}{\epsilon \tilde{\ell}} &=& V_{2S-n}^{\sigma \sigma^{'}}/\ \frac{\
e^2}{\epsilon \ell}, \label{eqn:fat}
\end{eqnarray}
where $\sigma$ and $\sigma^{\prime}$ (\,$= \, 1, 2 $\,) are layer indices,
while $V_{2S-n}^{11}=\,V_{2S-n}^{22}$ and $V_{2S-n}^{12}\,=\,V_{2S-n}^{21}$
are the intra- and inter-layer pseudopotentials for the relative angular
momentum $n=2S-J$, respectively \cite{qhehal}.

We can now plug this transformation into the $\nu=1$ SMA formula for the
pseudospin-wave mode, which is expressed in terms of $6j$ symbols, $\{^{S S
L}_{S S J}\}$, in the spherical geometry \cite{issp}.
Then we arrive at the desired expression for the pseudospin-wave spectrum,
$\omega _L$, for $\nu=1/m$ as
\begin{eqnarray}
\omega _L &=& \sqrt{e_L\,(e_L + 2 \lambda _L)}\,, \nonumber \\
e_L &=& \Delta _{\rm SAS} + \sum _{J=0}^{2\tilde{S}}
(2J+1)\,(-1)^{2\tilde{S}-J}\,\tilde{V}_J^{12}
\biggl[\,\frac {1}{2\tilde{S}+1}\,-\,(-1)^{2\tilde{S}-J}\, \left\{
\begin{array}{ccc} \tilde{S} & \tilde{S} & L\\
\tilde{S} & \tilde{S} & J \end{array} \right\}\,\biggl] , \nonumber \\
\lambda _L &=& \sum _J^{2\tilde{S}-J: \,{\rm odd}}
(2J+1)\,(\tilde{V}_J^{11} - \tilde{V}_J^{12})\,(-1)^{2\tilde{S}-J}
\left\{ \begin{array}{ccc} \tilde{S} & \tilde{S} & L\\ \tilde{S} &
\tilde{S} & J \end{array} \right\}, \label{eqn:cfsma}
\end{eqnarray}
where the range of the total angular momentum now reduces to $0\leq L \leq
2\tilde{S}$.
Since the pseudopotentials are shifted to a higher side of the relative
angular momentum (\,where the potential is softer\,), this approach could
be called the pseudopotential-shifted single-mode approximation
\cite{comment}.
We can confirm that, in the absence of tunneling (\,$\Delta _{\rm
SAS}=0$\,), the above formula reduces to a gapless Goldstone mode with
$\{^{S S \,\mbox{{\footnotesize 0}}} _{S S J} \} = (-1)^{2S-J}/ (2S+1)$. As
for the tunneling, we can regard that the flux attachment (\,or the
singular gauge transformation\,) does not affect the vertical interlayer
tunneling.

We now compare the composite-fermion SMA result with numerical ones for
finite systems.
Figure \ref{fig1} shows the low-lying excitation spectrum at $\nu=1/3$ for
$d/\ell =1.0$ and $\Delta _{\rm SAS}/(e^2/\epsilon \ell) = 0.01$ (a) or
$0.05$ (b).
For the numerical diagonalization we take a $5$-electron system, the
largest size tractable in the presence of tunneling \cite{odd}. The
charging energy is naturally included in the diagonalization.

We can see that the pseudospin-wave mode in the finite system does indeed
appear in the truncated range $0 \leq L \leq 2\tilde{S}=N-1$, while naively
there is no reason why the states should not extend for $0 \leq L \leq 2S =
m(N-1)$.
As for the dispersion curve itself, the CF prediction,
Eq.(\ref{eqn:cfsma}), exhibits a good agreement with the exact result up to
the wavenumber $k \sim \ell ^{-1}$.
This is the case with both Fig.\ref{fig1}(a), where a precursor of the
softening of the pseudospin-wave mode is visible, and Fig.\ref{fig1}(b)
with one-component character (\,see below\,). Thus the message here is that
the $\nu=1/m$ double-layer FQH system may be mimicked by a system of
composite fermions at $\nu=1$ with effectively {\em reduced}
\,inter-particle interactions for pseudospin-waves even in the presence of
tunneling.
It may be interesting if these collective modes are experimentally observed
e.g. with grated samples \cite{sohn}.

If we have a closer look at Fig.\ref{fig1}(b), we can see that $\Delta_{\rm
SAS}/(e^2/\epsilon \ell) = 0.05$ is large enough to make the ground state
almost a $\nu =1/3$ Laughlin liquid within the symmetric band with little
mixing of the antisymmetric band.
This is signaled by the fact that the lowest-lying excitation is a
magnetoroton mode within the symmetric band (\,identified by its
pseudospin\,), where the intra-band roton costs less energy than the
inter-band pseudospin wave.
The increased importance of $\Delta _{\rm SAS}$ relative to the
interparticle interaction is responsible for the situation.

The increased importance of $\Delta _{\rm SAS}$ also appears in the phase
diagram on the $\Delta_{\rm SAS}-d$ plane for the $\nu=1/m$ double-layer
FQH state.
The result (\,Fig.\ref{fig2}\,), obtained by identifying the softening of
the pseudospin-wave mode as the disappearance of the QHE gap, shows that
the QHE region widens as $\nu$ becomes $1/3, 1/5, \ldots$. This is again
because the electron-electron interaction is effectively weakened as we
attach two, four, $\ldots$ fluxes, which in turn reduces the mixing of the
antisymmetric state in the ground state to push the system toward the
one-component FQH state.
Thus the CF picture gives a natural explanation of the persistence of the
one-component character of the $\nu=1/3$ FQH state observed in a wide
single quantum well \cite{suen}, where $\Delta _{\rm SAS}$ is intrinsically
large, while the reduction of the interaction due to the expansion of the
wavefunction in wide wells, evoked in Ref.\cite{suen}, will be
quantitatively a secondary contribution.

Now we comment on the possibility of the `Josephson-like' interlayer
current in the $\nu=1/m$ double-layer system as discussed by several
authors \cite{bcsdl,wen}.
Since this effect should be related to the transition between Halperin's
$\Psi _{mmm}^{N_1}$ states [\,$N_1$\,($=0,1,\cdots,N$): the number of
electrons in layer 1\,], which are the $k=0$ states, an attention should be
paid on the gap at $k=0$.
When the interlayer transfer of electrons occurs over the whole
two-dimensional area, a finite gap emerges at $k=0$ [\,explicitly, $\Delta
_0 \equiv \sqrt{\Delta _{\rm SAS}(\Delta _{\rm SAS}+2 \lambda _0)}$ from
Eq.(\ref{eqn:cfsma})\,], and this will,
as pointed out by MacDonald and Zhang \cite{sma}, refute the dc
Josephson-like effect.
The suppression of the `Josephson effect' is also discussed in the
side-by-side geometry by Feng {\em et al} \cite{feng}, who pointed out that
a composite particle leaves behind phase fluctuations of the CS field when
it tunnels laterally.
By contrast, in a vertical tunneling in a double-layer geometry, a
composite particle can readily tunnel since it does not have to shake off
the flux quanta as seen in the present result.
Therefore, the \lq Josephson effect' may still be possible in a geometry
where the two layers have only a weak (\,e.g., spatially-localized\,)
vertical tunneling that may be regarded as a perturbation, while the
inter-layer electron correlation over the whole area continues to `lock'
the CS phase.

We are grateful to Drs K. Kusakabe and P. Maksym for valuable discussions.
The numerical calculations were done on HITAC S3800 in the Computer Center,
the University of Tokyo. This work was in part supported by a Grant-in-Aid
from the Ministry of Education, Science and Culture, Japan.



\begin{figure}
\caption{ The excitation spectrum from the ground state (\,origin of the
figure\,) for a double-layer $5$-electron FQH system at $\nu=1/3$, which
comprises pseudospin-wave excitations (\,open circles\,) and other
excitations (\,solid circles\,), is shown for $d/\ell =1.0$, and $\Delta
_{\rm SAS}/(e^2/\epsilon \ell) = 0.01$ (a) or $0.05$ (b). The results for
the pseudospin-wave excitation in the composite-fermion SMA for the same
number of electrons (\,crosses\,) and for a $51$-electron system (\,solid
line\,) are also shown. In (b), the roton excitations within the symmetric
band are also displayed by open squares. The roton and pseudospin-wave
excitations are respectively connected by curves as a guide to the eye.}
\label{fig1}
\end{figure}
\begin{figure}
\caption{The composite-fermion SMA result for the phase diagram for the
$\nu=1/m$ pseudospin-polarized
quantum-Hall state [\,the region below
solid line ($\nu=1$), broken line ($\nu=1/3$), or dotted line
($\nu=1/5$)\,].} \label{fig2}
\end{figure}


\begin{references}
\bibitem{comment} In the conventional SMA \cite{sma}, the correlation
effect is plugged into the the radial distribution function of the ground
state, which is assumed to be Halperin's $\Psi _{mmn}$,
while in the present CF approach we can continue to use the analytic
formula for $\nu=1$.
\bibitem{odd} We have checked that odd or even number of electrons does not
significantly alter the result, which may be due to the pseudospin
polarization.
\end{references}
\end{document}